# High-Capacity Reversible Data Hiding in Encrypted Images using Adaptive Encoding

MA Wen-Jing[1,2],   WU You-Qing[1],   YIN Zhao-Xia[1]

[1](Anhui Provincial Key Laboratory of Multimodal Cognitive Computation, School of Computer Science and Technology, Anhui University, Hefei 230601, China)

[2](School of Computer Science and Technology, Hefei Normal University, Hefei 230601, China)

Corresponding author: Zhaoxia Yin. E-mail: yinzhaoxia@ahu.edu.cn

**Abstract**:   With the popularization of digital information technology, the reversible data hiding in encrypted images (RDHEI) has gradually become the research hotspot of privacy protection in cloud storage. As a technology which can embed additional information in encrypted domain, extract the embedded information correctly and recover the original image without loss, RDHEI has been widely paid attention by researchers. To embed sufficient additional information in the encrypted image, a high-capacity RDHEI method using adaptive encoding is proposed in this paper. Firstly, the occurrence frequency of different prediction errors of the original image is calculated and the corresponding adaptive Huffman coding is generated. Then, the original image is encrypted with stream cipher and the encrypted pixels are marked with different Huffman codewords according to the prediction errors. Finally, additional information is embedded in the reserved room of marked pixels by bit substitution. The experimental results show that the proposed algorithm can extract the embedded information correctly and recover the original image losslessly. Compared with similar algorithms, the proposed algorithm makes full use of the characteristics of the image itself and greatly improves the embedding rate of the image. On UCID, BOSSBase, and BOWS-2 datasets, the average embedding rate of the proposed algorithm reaches 3.162 bpp, 3.917 bpp, and 3.775 bpp, which is higher than the state-of-the-art algorithm of 0.263 bpp, 0.292 bpp, and 0.280 bpp, respectively.

In the past few decades, with the continuous development and improvement of digital media technology, technologies for digital media protection have emerged in an endless stream [1],[2]. As an important information protection technology, information hiding plays a vital role in information security by realizing the functions of copyright protection, concealment communication, and so on, while not excessively affecting carriers. According to different technical objectives, traditional information hiding can be divided into three categories, namely digital watermarking, steganography and steganalysis, and reversible data hiding (RDH). Digital watermarking[3],[4] embeds information without affecting the usage of the original digital works. It is widely used in security fields such as copyright protection and integrity authentication.Robustness is an important attribute of digital watermarking, indicating the ability to extract and verify watermark information after various attacks, such as removal attacks and geometric attacks. Steganography [5-7] implements covert communication with digital media content as the carrier, focusing on imperceptibility, that is, steganography behavior can neither be perceived by human eyes nor detected by various steganographic analysis algorithms [1],[8]. Different from digital watermarking and steganography, early RDH[9],[10] is mainly applied in the plaintext domain. The embedded information can be extracted correctly and the original image can be recovered losslessly in plaintext domain RDH. Rate-distortion performance [11] is the common evaluation index of the three, which requires image distortion to be reduced as much as possible under the condition of obtaining the same embedding rate.

Reversible data hiding in plaintext domain is constantly developing and improving. The existing methods mainly include lossless compression [12],[13], difference expansion [14,15] and histogram shifting[9],[16], which greatly improve the performance of reversible information hiding. However, the reversible information hiding in plaintext domain obtains the image which is similar to the original carrier after embedding the information and the image content may be leaked, which is not allowed in some sensitive fields.For example, the disclosure of patient privacy in medical diagnosis can cause inconvenience to patients, and the disclosure of state secrets in military images can cause irreparable situations. Image encryption can protect personal privacy   by encrypting the carrier signal and make it difficult for illegal operators to obtain valuable information. Based on this, it is proposed to combine image encryption with information hiding, that is, the image owner encrypts the original image, the information hider embeds the information in the encrypted image, and finally the legal receiver extracts the information or recovers the image. This Reversible Data Hiding in Encrypted Images(RDHEI) method [17-19] has attracted widespread attentions in recent years. This method has good application prospects for privacy protection in cloud storage, and can also be applied to medical and military image transmission and storage, as well as forensic evidence and other scenarios.

According to the sequence of image encryption and room reservation, the existing RDHEI methods can be divided into two main categories: vacating room after encryption (VRAE) RDHEI and reserving room before encryption (RRBE) RDHEI. The VRAE method [20-23] utilizes the image redundancy after encryption to vacate the room for embedding information. An RDHEI method utilized the spatial correlation of image is proposed in [20]. By flipping the least significant bits of the pixels in the encrypted image block, additional information can be embedded and extracted. By using the pixel correlation of natural image, this method can extract the embedded information successfully and recover the original image well. However, when the encrypted image block is too small, the extraction of embedded information maybe fail in the rough area of the image. Subsequently, a separable RDHEI method is proposed in [22]. The method vacates room to accommodate the embedded information by compressing the least significant bits of the encrypted image. At the image receiving end, the extraction of embedded information and the recovery of the image can be performed separably. This separable method further broadens the application scenarios of RDHEI. In [23], a new RDHEI framework using block encryption is proposed. In this method, the image is encrypted by blocks, and pixel correlation is still retained in the blocks after encryption, so most plaintext reversible data hiding methods can be used to vacate room in the encrypted image.This method also further proves that preserving the correlation between image pixels is crucial to vacate room for information embedding.

Above mentioned methods can embed additional information in the encrypted image, but due to the low image redundancy after encryption, the embedding capacity of the RRBE method is limited and some bit errors probably exist when recovering the image. Different from the VRAE method, the method of RRBE [24-27] processes the image before encryption to reserve room. Literature [24] first proposes an RDHEI method that reserves room before image encryption. This method realizes the correct extraction of embedded information and the lossless recovery of the image. Subsequently, an RDHEI method using the prediction error histogram shifting is proposed in [25]. According to different application requirements, the embedded information can be extracted from the encrypted image and the decrypted images respectively. Then, the most significant bit (MSB) of the pixel is predicted to reserve the room for the first time in [26]. Since the RDHEI method regardless of the loss of image quality in the encrypted domain, it is feasible to select the MSB of the pixel for prediction. Besides, this method further improves the embedding capacity of the image. Inspired by the MSB prediction method, some RDHEI methods based on multi-MSB prediction are proposed. Literature [27] introduces an algorithm based on multi-MSB prediction and Huffman coding. In this method, the binary sequence of the original pixel value and the prediction value is compared and the number of the same bits from MSB to LSB is recorded to generate Huffman coding. Then, each pixel is marked with the corresponding Huffman codeword. This method considers not only multi-MSB prediction but also pixel marking. The method of marking pixels receives attention from researchers recently.

In [28], an RDHEI method based on the parametric binary tree labeling is proposed. In the method, information can be embedded in the unmarked bits of the pixel by binary tree coding. Next, the parametric binary tree labeling method is extended in [29]. By reducing the impact of image blocks on pixel utilization, a higher embedding capacity can be obtained. Subsequently, Literature [30] proposes a method based on bit plane compression. This method not only recovers the original image losslessly but also greatly improves the embedding capacity. Compared with RDHEI method which reserve room before image encryption, processing prediction error before encryption can obtain higher embedding capacity.This is because the spatial redundancy of the original image is limited, and the fluctuation range of the prediction error corresponding to the image is relatively small, so more room can be reserved.

The method proposed in [29] utilizes equal length encoding to mark pixels with different prediction errors. This method can obtain a higher embedding capacity, but it does not consider the distribution characteristics of the prediction errors. Based on this, we propose a high-capacity adaptive encoding method in this paper to improve the method in [29]. In the proposed method, the pixels with different prediction errors are marked with adaptive variable-length Huffman coding, we also utilize the prediction error with high redundancy as the carrier to reserve more room. In the proposed method, first, the prediction error of the whole image is calculated and Huffman coding is generated according to occurrence probability of prediction error. Then, the generated Huffman codewords are used to mark the encrypted pixels by bit substitution and the remaining bits are used to embed additional information. Compared with the method in [29], the proposed method uses adaptive Huffman coding to mark pixels. By combing the texture characteristics of each image, the proposed method finally obtains a higher embedding capacity.

# 1  Research framework

In the RDHEI method, there are mainly three roles: the image owner, the information hider, and the image receiver. The image owner possesses the information of the original image, he performs a series of preprocessing and encrypts the image to protect the image content; The information hider cannot obtain the original image information without the permission of the image owner, but he can embed some necessary additional information in the encrypted image; The image receiver can execute the information extraction or image recovery according to the type of encryption key.

The texture characteristics of different images are distinct, so the corresponding prediction error distributions also vary greatly. To take advantage of the characteristic of each image, we propose a high-capacity RDHEI method using adaptive encoding in this paper. As shown in Fig.1, the proposed method can be divided into the following steps:

(1) The image owner preprocesses and encrypts the original image. The image owner first preprocesses the original image. On the one hand, the image owner calculates the prediction error of the image and generates Huffman coding according to the occurrence frequency of prediction error. On the other hand, the original image is directly encrypted to obtain the encrypted image. Then, the encrypted pixels are marked by bit substitution according to the Huffman codewords corresponding to prediction errors. Finally, the marked encrypted image is obtained;

(2) The information hider embeds additional information into the marked encrypted image. The unmarked bits of each pixel in the marked encrypted image are the reserved room. The information hider can embed the encrypted additional information into the reserved room by bit substitution. Although the information hider cannot obtain the information of the original image based on the marked encrypted image, it does not affect the embedding of the additional information and the generating of the stego image;

(3) The image receiver extracts the embedded information or recovers the original image. After the image receiver obtains the stego image, he performs the information extraction or image recovery operation according to different keys.

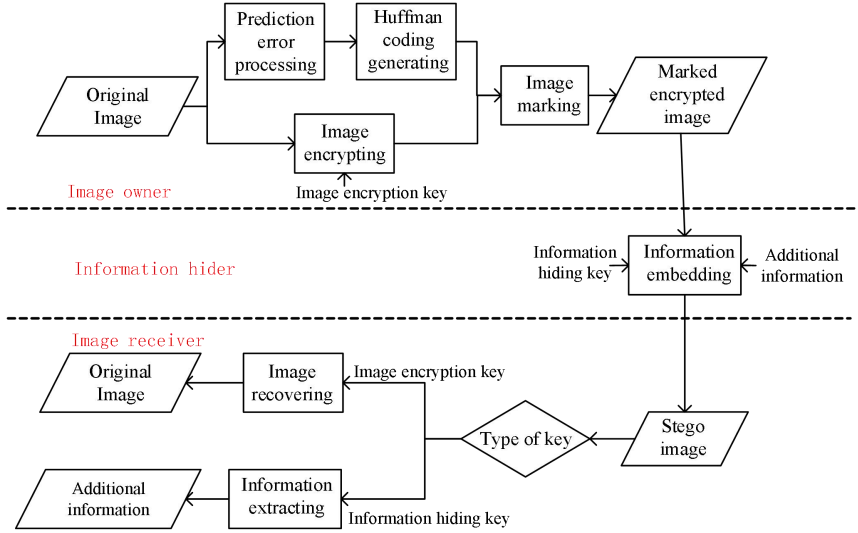

Fig.1　The framework of the proposed RDHEI method

## 2　High-capacity RDHEI using adaptive encoding

Section 1 introduces the general framework of the high-capacity RDHEI method using adaptive encoding. This section describes the details of the proposed method. In the first stage, the image owner calculates the prediction error of the image. The specific process is described in Section 2.1. Then in Section 2.2, the Huffman coding is adaptively generated according to the occurrence frequency of the prediction error. Next, the encrypted image is marked according to the Huffman codewords corresponding to different prediction errors to obtain the marked encrypted image. In the second stage, once the information hider receives the marked encrypted image, he embeds the encrypted additional information in the unmarked bits of the pixel to obtain the stego image. The detailed operation is introduced in Section 2.3. After receiving the stego image, the image receiver can perform information extraction or image recovery operations on the image, the details are introduced in Section 2.4.

### 2.1　Prediction error calculating

For a grayscale image with a size of m × n, the median edge detector(MED) method [31] can be used to calculate the prediction value of the original image. x(i,j) (2≤i≤m, 2≤j≤n) represents a pixel value of the original image. As shown in Fig.2, three pixels in the left, top, and top left of pixel x(i,j) are selected as reference values to calculate the prediction value p(i, j).

$$p(i,j) = \begin{cases} \max(X_2, X_3), X_1 \leq \min(X_2, X_3) \\ \min(X_2, X_3), X_1 \geq \max(X_2, X_3) \\ X_2 + X_3 - X_1, others \end{cases} \quad (1)$$

In the process of calculating the prediction values of the image, pixels in the first row and the first column are regarded as reference pixels without any operations. From the pixel in the second row and second column, the corresponding prediction value can be calculated according to formula (1). Then, the prediction error e(i,j) of pixel x(i,j) can be calculated by e(i,j)=x(i,j)-p(i,j). Finally, the prediction errors of the remaining pixels are also calculated in turn according to the above method, until the whole prediction error of the image is obtained. On the one hand, the MED method utilizes the neighboring pixels to calculate the prediction value, which can obtain more accurate prediction value of pixels with low complexity. On the other hand, the MED method is convenient for image recovery. Since the three pixels adjacent to the predicted pixels are reference values, and the pixels in the first row and the first column are retained, the image pixels can be scanned sequentially to recover the image.

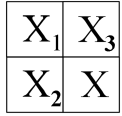

Fig.2　MED predictor is used for pixel prediction

**2.2　Huffman coding and marking**

To utilize the texture characteristic of the image and reserve room for embedding additional information, it is necessary to perform adaptive Huffman coding on the prediction error of the image. The generated Huffman coding is used to mark pixels, so the length of the Huffman codeword cannot exceed eight bits. To satisfy this restriction, the occurrence frequency of each prediction error should be preprocessed. Assuming that the length of each Huffman codeword is 8, the corresponding Huffman coding is the situation where the most codewords can be obtained. The average occurrence frequency of the corresponding prediction error at this time is recorded as the partition threshold to determine the range of the initial prediction error. Fig.3 shows the Huffman tree when all Huffman codewords are 8 bits. At this time, the average frequency of each codeword is 1/256, which is recorded as the initial partition threshold. The prediction error whose occurrence frequency exceeds the initial partition threshold is recorded as a case, and the sum of the occurrence frequency is calculated. The prediction errors with occurrence frequency within the initial partition threshold can be divided into s kinds of cases (s is determined by the kinds of prediction errors within the partition threshold). Finally, Huffman coding is generated according to the occurrence frequency of s+1 kinds of prediction error. Pixels with lowly occurrence frequencies of prediction errors are recorded as one case and the sum of these occurrence frequencies is counted. The occurrence frequency of the prediction errors of other pixels are also recorded one by one.　According to the initial partition threshold, the range of the initial prediction error is determined and corresponding Huffman coding is carried out.

Determine whether the current Huffman codewords satisfy the restriction. If not, change s to s-2 until the prediction error range meets the restriction. Then, the final partition threshold is obtained. In the end, we obtain Huffman coding that meets the characteristic of the image texture.

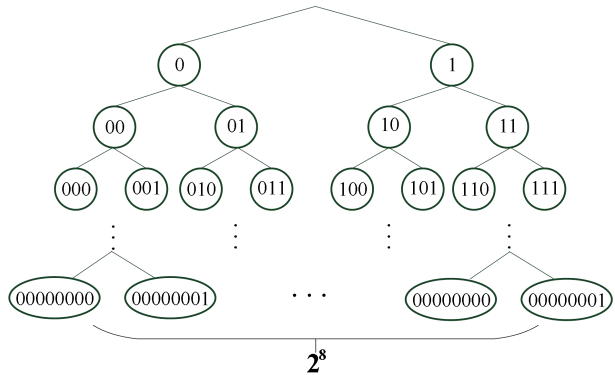

Fig.3　A Huffman tree corresponding to a Huffman encoding length of 8

In the process of generating Huffman coding, pixels can be divided into three types. They are reference pixels, non-embedded pixels, and embeddable pixels. The reference pixels are located in the first row and the first column of the image without any operations. The rest pixels are divided into non-embedded pixels and embeddable pixels according to the final partition threshold. The pixels whose occurrence frequency of prediction errors exceed the final partition threshold are recorded as non-embedded pixels, otherwise, they are embeddable pixels.

To protect the information of the image from being leaked, the image encryption should be performed before marking the pixels with Huffman codewords. In image encryption, a random matrix H with the same size as the image is generated according to the image encryption key, h(i,j) is the value in H, i and j represent abscissa and ordinate of the corresponding value. The original pixel value x(i,j) and h(i,j) are converted into eight-bit binary number, the specific conversion operation is shown in formula (2), floor(*) is the floor operation, mod represents the complementary function and k is the number of bits corresponding to the binary number. After converting x(i,j) and h(i, j) to corresponding eight-bit binary number, the encryption is performed as formula (3), '⊕' represents the exclusive-or operation. Then the eight-bit binary number after the exclusive-or encryption is converted to a decimal number according to the formula (4). That is, the encrypted pixel value $x_e(i,j)$. Finally, the above operations are performed on all pixels in the image to obtain the encrypted image.

$$y^k(i,j) = \left[floor\left(\frac{y(i,j)}{2^{k-1}}\right)\right] mod\ 2,\ k = 1,2,...,8 \quad (2)$$

$$x_e^k(i,j) = x^k(i,j) \oplus h^k(i,j),\ k = 1,2,...,8 \quad (3)$$

$$x_e(i,j) = \sum_{k=1}^{8} x_e^k(i,j) \times 2^{k-1},\ k = 1,2,...,8 \quad (4)$$

After the image is encrypted, the encrypted pixels can be marked with Huffman codewords by bit substitution. According to the different types of pixels, the operations when marking pixels are also distinct. The reference pixels are not marked and their information is recorded as auxiliary information. Non-embedded pixels are marked with the same Huffman codeword and the information of the replaced non-embedded pixels is also recorded as auxiliary information. Embeddable pixels are marked with different Huffman codewords according to different prediction errors. The unmarked bits of embeddable pixels are the reserved room and can be used to embed additional information. Once the pixels are marked, the Huffman coding rules are converted into a binary bitstream and stored in the position of the original reference pixels by bit substitution. To facilitate subsequent information embedding or extracting, the auxiliary information is embedded in the reserved room of the current embeddable pixels by bit substitution.

| 162 | 162 | 162 | 161 |
|---|---|---|---|
| 162 | 162 | 162 | 161 |
| 162 | 162 | 162 | 161 |
| 162 | 162 | 162 | 161 |

(a) Original image

| 162 | 162 | 162 | 161 |
|---|---|---|---|
| 162 | 0 | 0 | 0 |
| 162 | 0 | 0 | 0 |
| 162 | 0 | 0 | 0 |

(b) Prediction error

| 00100001(33) | 11100010(226) | 10010010(146) | 10101101(173) |
|---|---|---|---|
| 11010111(215) | 01111010(122) | 10001110(142) | 10011101(157) |
| 11111011(251) | 00101000(40) | 01001001(73) | 00101100(44) |
| 10111010(186) | 10100001(161) | 00100100(36) | 00010001(17) |

(c) Encrypted image

| 00100001(33) | 11100010(226) | 10010010(146) | 10101101(173) |
|---|---|---|---|
| 11010111(215) | 10011010(154) | 10001110(142) | 10011101(157) |
| 11111011(251) | 10001000(136) | 10001001(137) | 10001100(140) |
| 10111010(186) | 10000001(129) | 10000100(132) | 10010001(145) |

(d) Marked the encrypted image with Huffman coding

| 10000100(132) | 01000111(71) | 01001001(73) | 10110101(181) |
| 11101011(235) | 01011001(89) | 01110001(113) | 10111001(185) |
| 11011111(223) | 00010001(17) | 10010001(145) | 00110001(49) |
| 01011101(93) | 10000001(129) | 00100001(33) | 10001001(137) |

(e) Marked encrypted image

Fig.4　Diagram of pixel marking process:(a) Original image, (b) Prediction error, (c) Encrypted image, (d) Marked the encrypted image with Huffman coding, (e) Marked encrypted image

　　Fig.4 captures a part of the Lena image and briefly explains the whole process of pixel marking. Fig.4(a) is the original image, the first row and the first column of pixels are reference pixels. According to the calculation method of the prediction error in Section 2.1, the prediction error is shown in Fig.4(b). The prediction error at this time is preprocessed and coded to obtain the Huffman codewords shown in Table 1. Fig.4(c) is the encrypted image obtained by directly encrypting Fig.4(a) with the image encryption key. Then, the encrypted image is marked by bit substitution according to the Huffman codewords shown in Table 1. As shown in Fig.4(d), the information in red is the Huffman marking bits corresponding to the current pixel and the remaining bits represent the reserved room, in which information can be embedded. As the prediction errors of adjacent pixels often have small differences, to prevent image information from being leaked, Fig.4(d) is reversed to obtain the marked encrypted image shown in Fig.4(e).

Table 1　Huffman coding table

| Prediction error | Huffman codewords |
|---|---|
| -5 | [0,1,0,0,1] |
| -4 | [1,1,0,1] |
| -3 | [0,1,1,0] |
| -2 | [0,0,1,0] |
| -1 | [0,0,0,0] |
| 0 | [1,0,0] |
| 1 | [1,1,1] |
| 2 | [0,0,0,1] |
| 3 | [0,1,0,1] |
| 4 | [1,1,0,0] |
| 5 | [0,1,0,0,0] |

**2.3　Embed additional information**

　　In the marked encrypted image, assuming that the length of the auxiliary information is $l$. There are $t$ embeddable pixels $\{x_1, x_2, x_3, ..., x_t\}$ in the marked encrypted image. Each embeddable pixel $x_p(1 \leq p \leq t)$ is marked with a Huffman codeword whose length is $n_p(n_p \leq 8)$. Then each embeddable pixel $x_p$ can reserve $8 - n_p$ bits for embedding information. The net embedding capacity $c$(bit) can be calculated based on the total number of the reserved room and the length of auxiliary information.

$$c = \sum_{p=1}^{t}(8 - n_p) - l \quad (5)$$

　　When the information hider obtains the marked encrypted image, the Huffman coding rules stored in the reference pixels can be obtained. Combined with Huffman coding rules, the position of the net reserved room can be located. Then, the encrypted additional information is embedded in the net reserved room by bit substitution to generate the stego image. To ensure that the content of the additional information is not leaked, the additional information must be encrypted by the information hiding key, the detailed encryption operation is the same as the image encryption method in Section 2.2.

**2.4　Information extraction or image recovery**

　　In the information extraction or image recovery stage, the image receiver extracts and obtains the Huffman coding rules from the reference pixels. According to the Huffman coding rules, the image receiver can find all embeddable pixels and extract the embedded information. The extracted information is mainly composed of two parts, which are auxiliary information and encrypted additional information. With different encryption keys the image receiver possesses, the information extraction or image recovery can be processed separately. Specific operations can be divided into the following three cases:

(1) Only the image encryption key: When the image receiver only has the image encryption key, the original image can be recovered losslessly. The image receiver recovers the extracted auxiliary information to the corresponding position to obtain the encrypted reference pixels and non-embedded pixels. With the image encryption key, the original reference pixels and the original non-embedded pixels of the image can be recovered. Then, different prediction errors of pixels can be obtained by Huffman coding rules. Finally, the image receiver calculates the prediction value of each pixel according to formula (1) and recovers the embeddable pixels with the prediction value and prediction error. After the above operations, the stego image is recovered to the original image.

(2) Only the information hiding key: When the image receiver only has the information hiding key, the embedded additional information can be extracted correctly. Then, the image receiver decrypts the extracted additional information with the information hiding key to recover it.

(3) Both the image encryption key and the information hiding key: When the image receiver has both the image encryption key and the information hiding key, the additional information can be extracted correctly or the original image can be recovered losslessly. In the process of information extraction and image recovery, they do not affect each other, so the information extraction and image recovery is separable.

## 3 Experimental results and analysis

Experiments are designed in this section to verify the feasibility of the proposed method. As shown in Fig.5, five conventional grayscale images are used for experimental comparison to verify the feasibility of the proposed method. At the same time, to reduce the impact of the texture complexity of the test images on the experiment, we also test the performance of different methods in three datasets of UCID[32], BOSSBase[33], and BOWS-2[34]. Because the RDHEI method does not care about the image quality in the encrypted domain, the embedding performance has become an important indicator. The embedding rate (ER) is regarded as the key indicator to measure the embedding performance and the ER is expressed in bpp (bit per pixel). The maximum value of the embedded additional information is selected to calculate ER. Besides, we also use common indicators Mean Square Error (MSE) and Structural Similarity (SSIM) to test the reversibility of the proposed method.

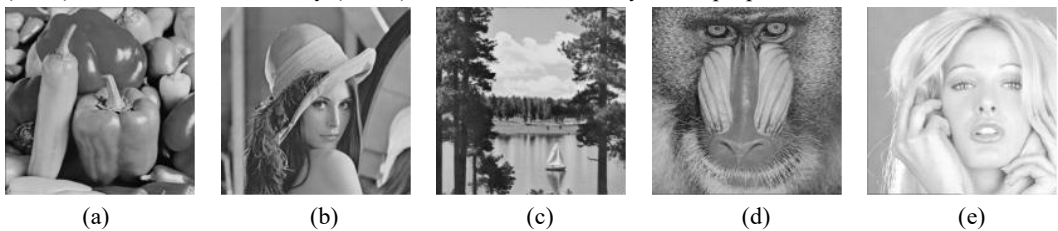

(a)　　　　　　　　(b)　　　　　　　　(c)　　　　　　　　(d)　　　　　　　　(e)

Fig.5　Five grayscale test images: (a) Peppers, (b) Lena, (c) Lake, (d) Baboon, (e) Tiffany

### 3.1　Reversibility analysis

To verify that the proposed method can extract the information or recover the image reversibly, this section carries out a verification of reversibility. In the information extraction or image recovery stage, the prediction errors can be obtained and the original image can be recovered losslessly. Three datasets of UCID[32], BOSSBase[33], and BOWS-2[34] are utilized to verify the reversibility of the image recovery. Table 2 shows the MSE and the SSIM of original datasets and recovered datasets. The MSE of the datasets is '0', indicating that the recovered image is consistent with the original image. At the same time, the SSIM of the recovered image and the original image is '1', which means that the structure of the two images is uniform. The experimental result in Table 2 proves that the proposed method can recover the original image reversibly. The information extraction of the proposed method is introduced in Section 2.4. In the process of information extraction, the image receiver extracts all the embedded additional information losslessly, indicating that the proposed method can also achieve the reversibility of information extraction. In the proposed method, the process of information extraction and image recovery does not affect each other and can be completed independently. The above analyses further prove that the information extraction or image recovery of the proposed method is not only reversible but also separable.

Table 2    MSE and SSIM of original datasets and recovered datasets

|      | UCID | BOSSBase | BOWS-2 |
|------|------|----------|--------|
| MSE  | 0    | 0        | 0      |
| SSIM | 1    | 1        | 1      |

## 3.2    Safety analysis

As a high-capacity RDHEI method using adaptive encoding, the proposed method not only protects the original image content from being leaked but also protects the additional information embedded in the image. To prove the security of the proposed method, the encryption method and the feature map of images in different states are analyzed in this section. In the above introductions, the image encryption key and the information hiding key are utilized in the proposed method for security. For a grayscale image of size $m \times n$, the image encryption key generates a random matrix H with the same size as the image. After each pixel in H converting into an eight-bit binary number, there are $8 \times m \times n$ bits '0' or '1'. So the existence possibility of the image encryption key is $2^{8 \times m \times n}$, which is very difficult to predict the correct image encryption key. Similarly, assuming that g bits of additional information are embedded in the marked encrypted image, there are $2^g$ kinds of existence possibility of the information hiding key, it is also difficult to obtain the correct information hiding key. The existence of the image encryption key and the information hiding key makes it difficult for illegal users to obtain the plaintext information and the embedded additional information of the image. Besides, the pixel characteristic of the image at each state in Fig.6 further confirms the security of the proposed method. Fig.6(a) is the original Lena image and Fig.6(b) is the encrypted image. Then the Huffman coding is utilized to obtain the marked encrypted image as shown in Fig.6(c). After embedding additional information, the stego image Fig.6(d) is obtained. Finally, image recovery is performed to obtain the recovered image Fig.6(e). Fig.6(f) shows the histogram of pixel distribution of the original image and Fig.6(g) is the histogram of pixel distribution of the encrypted image. The pixel distribution of the image in Fig.6(g) is flat, indicates that meaningful image information is hard to be obtained. After the encrypted image is marked by Huffman coding, the marked encrypted image is obtained. Fig.6(g) is the histogram of pixel distribution of the marked encrypted image. Compared with the histogram of pixel distribution of the encrypted image, the overall distribution of the pixel in Fig.6(g) is still messy, which means that the proposed RDHEI method is guaranteed in terms of security.

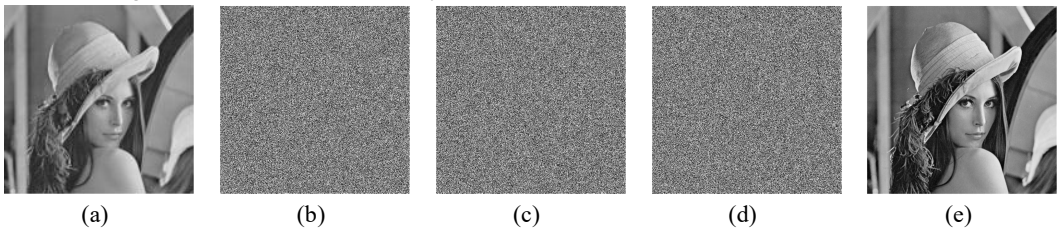

(a)            (b)            (c)            (d)            (e)

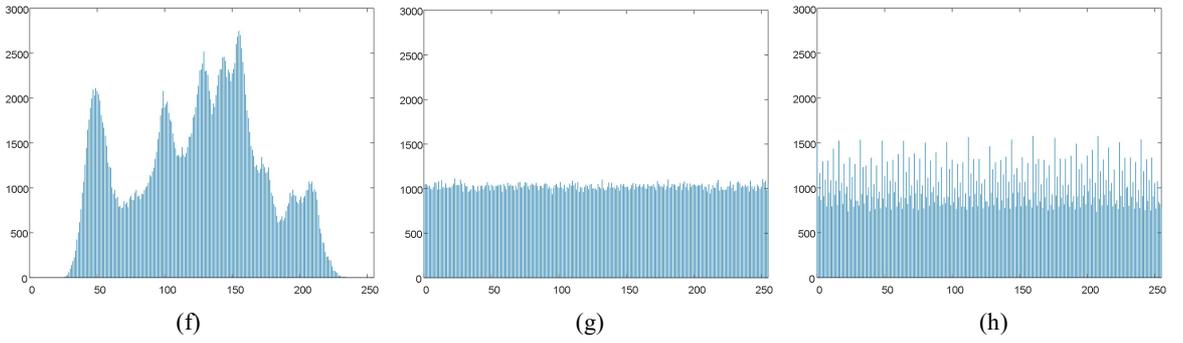

Fig.6   Image pixel feature representation schematic diagram of different stages: (a) Original image, (b) Encrypted image, (c) Marked encrypted image,(d) Stego image, (e) Recovery image, (f) Histogram of the original image, (g) Histogram of the encrypted image, (h) Histogram of the marked encrypted image

### 3.3   Performance comparison

The proposed method utilizes the adaptive Huffman coding to mark pixels to improve the method in [29], which makes full use of the texture characteristic of the image. The unmarked bits of each embeddable pixel can be used to embed information. Combined with Formula 5, the embedding rate can be obtained with $ER = c/(m \times n)$ (bpp). Table 3 shows the performance comparisons between [29] and the proposed method from many aspects. As shown in Table 3, the number of embeddable pixels is increased and the number of auxiliary pixels (reference pixels and non-embedded pixels) that cannot be used to reserve room is reduced in the proposed method. Compared with the method in [29], the pixel utilization is increased to a certain extent so that ER is significantly improved in the proposed method. Since the proposed method utilizes adaptive Huffman coding to mark the image, pixels corresponding to the prediction errors with higher occurrence frequency are marked with shorter Huffman codewords, which finally makes full use of the characteristics of the image itself and reserves more room.

Also, to more fully illustrate the performance of the proposed method, the ER in the proposed method is compared with the method in [26], [28], [29], and [30]. The method in [26] predicts the MSB to embed information and the ER is close to 1bpp. A parameter binary tree labeling method is utilized to mark image pixels in [28], which improves the embedding performance of the image. Then, the method in [29] uses an improved binary tree coding with equal length to mark more embeddable pixels, which greatly improves the ER. Combined with the characteristic of the prediction error, the ER in [30] is further improved by compressing the bit-plane of prediction error. As shown in Fig.7, the ER of the five test images in Fig.5 is compared firstly. The ER of the Baboon image in all methods is significantly lower than that of the other images. This is because the image texture of Baboon is more complex and the image redundant is low. Table 3 shows there are 36614 auxiliary pixels in the Baboon image, which is more than other test images. But the experimental results also show that the ER of the Baboon image obtained by the proposed method is still higher than other methods. For other test images, the proposed method can still obtain a higher ER .

To better verify that this performance improvement is not accidental, this section also conducts comparative experiments on the UCID[32], BOSSBase[33], and BOWS-2[34] datasets. The average ER in three datasets of the proposed method is compared with the method in [26], [28], [29], and [30] to verify the effectiveness of the proposed method. As shown in Fig.8, the proposed method can obtain a higher average ER on three datasets. The experimental results show that in the proposed method, the average ER on the datasets of UCID[32], BOSSBase[33], and BOWS-2[34] respectively reached 3.162bpp, 3.917bpp, and, 3.775bpp. Compared with other methods, the average embedding rate of the proposed method is improved significantly. Even compared with the literature [30] with the best performance, the average embedding rate of the proposed algorithm is still improved by

0.263bpp, 0.292bpp and 0.280bpp. Compared with other methods, the proposed method utilizes adaptive Huffman coding to mark pixels with different prediction errors. Combined the feature with the shortest average length of Huffman coding, the proposed method make full use of the texture characteristics of the image, which promotes the higher average ER.

Table 3  The performance comparison with TMM2020[29]

| Test Images | Method | Embeddable pixel | Auxiliary pixel | Pixel utilization=number of embeddable pixels/number of image pixels | ER(bpp) |
| --- | --- | --- | --- | --- | --- |
| Lena | TMM2020[29] | 243107 | 19037 | 92.7% | 2.645 |
| | Proposed method | 246614 | 15530 | 94.1% | 3.262 |
| Baboon | TMM2020[29] | 155266 | 106878 | 59.2% | 0.969 |
| | Proposed method | 225530 | 36614 | 86.0% | 1.481 |
| Tiffany | TMM2020[29] | 243463 | 18681 | 92.9% | 2.652 |
| | Proposed method | 245814 | 16330 | 93.8% | 3.376 |
| Peppers | TMM2020[29] | 235184 | 26960 | 89.7% | 2.494 |
| | Proposed method | 249882 | 12262 | 95.3% | 2.910 |
| Lake | TMM2020[29] | 209185 | 52959 | 79.8% | 1.998 |
| | Proposed method | 243183 | 18961 | 92.8% | 2.409 |

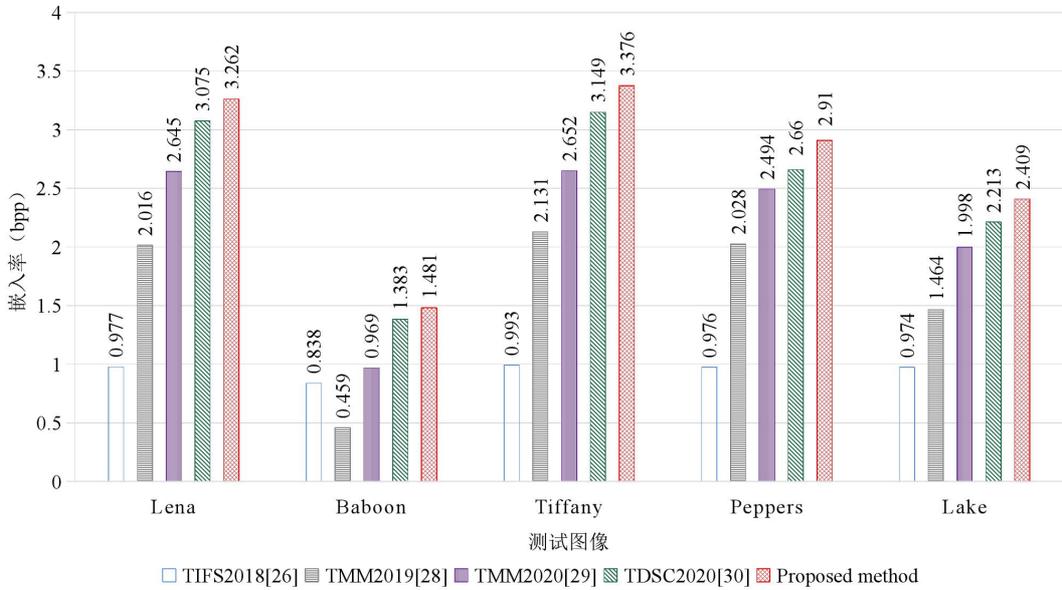

Fig.7  Comparison of ER (bpp) of five test images between our method and state-of-the-art methods

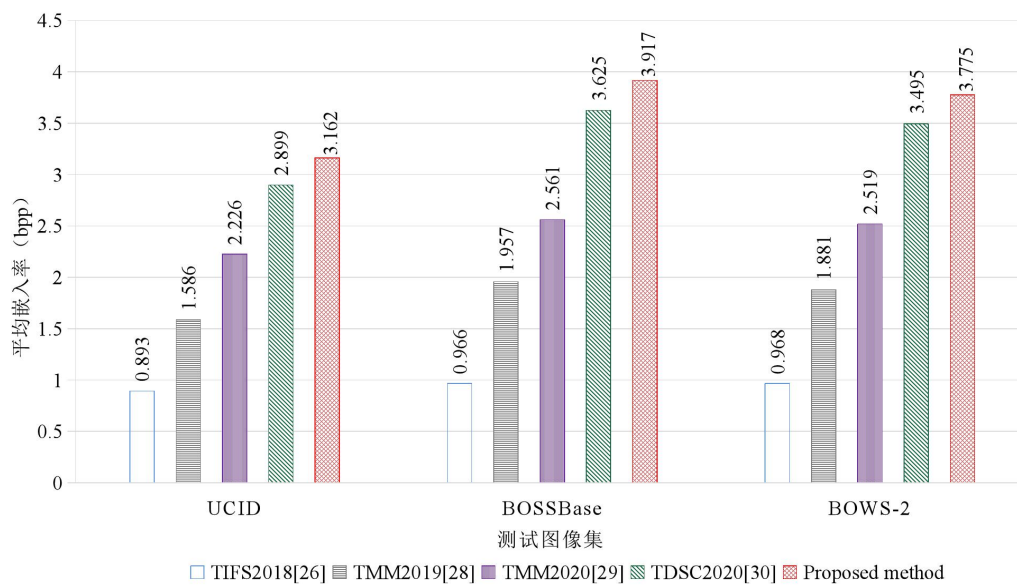

Fig.8　Comparison of the average ER(bpp) of three datasets between our method and state-of-the-art methods

## 4　Summary

　　The reversible data hiding plays an important role in privacy protection, and embedding performance is an important index to measure this method. This paper improves the method in [29] and proposes a high-capacity RDHEI method using adaptive encoding. According to the occurrence frequency of different prediction errors of the image, the proposed method utilizes adaptive Huffman coding to mark pixels. Compared with the equal length coding method proposed in [29], the proposed adaptive Huffman coding method is more flexible and the coding with the shortest average codewords can be obtained. At the same time, pixels can be marked according to the occurrence frequency of prediction error, so that more pixels can be marked and more embedding room can be reserved. Experimental results show that the proposed method can extract information correctly and recover images losslessly. Compared with similar methods, the proposed method can obtain higher embedding rate and provide sufficient room for embedding more additional information. However, the proposed method still has some non-embedded pixels in pixel marking. In the future, we can try to continuously improve the number of embeddable pixels, or explore new predictors to generate prediction errors with smaller fluctuation range and obtain more embeddable pixels, so as to further improve the embedding performance.